\documentclass[conference]{IEEEtran}
\IEEEoverridecommandlockouts
\pdfoutput=1
\usepackage{cite}
\usepackage{amsmath,amssymb,amsfonts}
\usepackage{algorithmic}
\usepackage{graphicx}
\usepackage{textcomp}
\usepackage{hyperref}

\usepackage{xcolor}
\usepackage{subcaption}
\def\BibTeX{{\rm B\kern-.05em{\sc i\kern-.025em b}\kern-.08em
    T\kern-.1667em\lower.7ex\hbox{E}\kern-.125emX}}
    
\DeclareCaptionFormat{myformat}{\fontsize{8}{8}\selectfont#1#2#3}
\usepackage[all]{nowidow} 
\usepackage[protrusion=true,expansion=true]{microtype} 
    
\begin{document}

\title{Dataset of Spatial Room Impulse Responses in a Variable Acoustics Room for Six Degrees-of-Freedom Rendering and Analysis\\
}

\author{\IEEEauthorblockN{Thomas McKenzie$^\star$, Leo McCormack, Christoph Hold}
\IEEEauthorblockA{\textit{Department of Signal Processing and Acoustics}, \textit{Aalto University}, Espoo, Finland \\
$^\star$thomas.mckenzie@aalto.fi}
}

\maketitle

\begin{abstract}
Room acoustics measurements are used in many areas of audio research, from physical acoustics modelling and speech enhancement to virtual reality applications. This paper documents the technical specifications and choices made in the measurement of a dataset of spatial room impulse responses (SRIRs) in a variable acoustics room. Two spherical microphone arrays are used: the mh Acoustics Eigenmike em32 and the Zylia ZM-1, capable of up to fourth- and third-order Ambisonic capture, respectively. The dataset consists of three source and seven receiver positions, repeated with five configurations of the room's acoustics with varying levels of reverberation. Possible applications of the dataset include six degrees-of-freedom (6DoF) analysis and rendering, SRIR interpolation methods, and spatial dereverberation techniques. 
\end{abstract}

\begin{IEEEkeywords}
Spatial room impulse response, six degrees-of-freedom, variable acoustics, higher-order Ambisonics
\end{IEEEkeywords}



\section{Introduction}

\begin{figure}
  \centerline{\includegraphics[width=\linewidth,trim={1.5cm 11cm 23cm 0.95cm},clip]{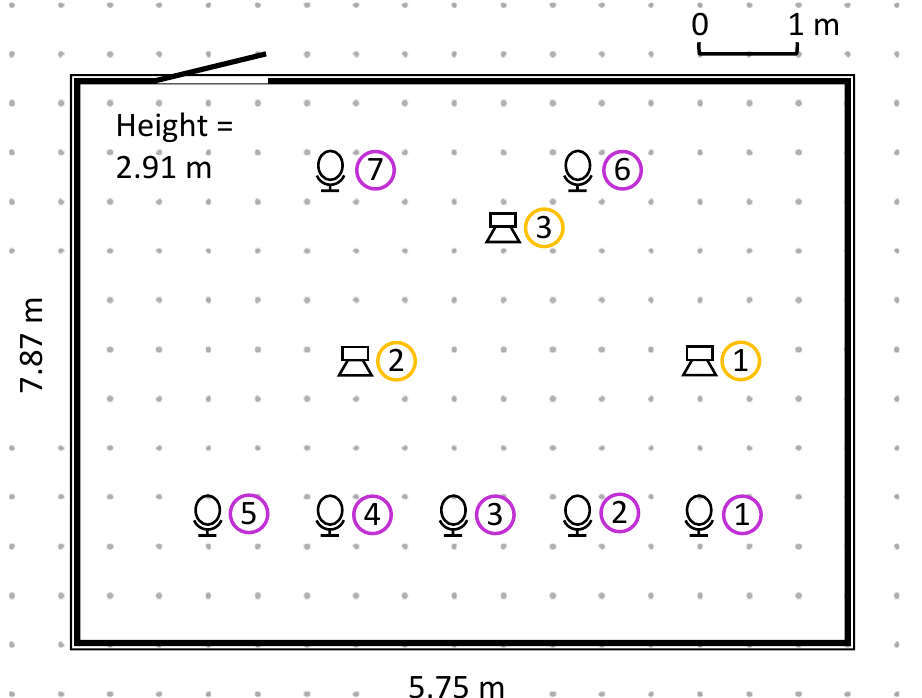}}
\caption{Illustration of the source and receiver positions. Microphones and microphones were oriented faced north and south, respectively, according to the illustration orientation. }
\label{fig:arni_roomplan}
\end{figure}

This paper details the acquisition of a dataset of SRIRs measured in the Arni variable acoustics room at the Acoustics Lab, Aalto University, Espoo, Finland -- consisting of three source positions, seven receiver positions and five different configurations of the room's acoustics. The source and receiver positions were selected to simulate a possible performance space or stage, whereby receivers were distributed closer to the walls and sources were placed in the centre of the room; with the sources all pointed towards the south wall (which could be interpreted as the `front' of the stage). Source and receiver positions are depicted in Fig.\,\ref{fig:arni_roomplan}, with exact coordinates presented in Table\,\ref{tab: coordinates}. The seven receiver positions were chosen as a line array of five receivers (R1 - R5), with an inter-receiver array distance of 1.25~m at a distance of 1.3~m from the south wall, and two near the north wall (R6 and R7). The three source positions were chosen as follows: S1 was placed slightly outside the geometry enclosed by the receiver positions (i.\,e.\ their planar convex hull), in order to produce a potentially challenging scenario for a reproduction workflow. S2 was placed within the convex hull of the receiver positions and was relatively central. S3 was placed nearer to the north wall, inside the convex hull of the receiver positions. All positions and orientations of the three sources and seven receivers were aligned to the best of the ability of the authors using laser level meters. However, small positional errors up to 2-3\,cm, and orientation errorsup to 1-2° are possible.

Loudspeakers were used as the sources, with their height (z-axis) measured from the floor to the centre of the coaxial driver. As according to Fig.\,\ref{fig:arni_roomplan} and Table\,\ref{tab: coordinates}, the x (counting west), y (counting north) coordinates of the loudspeakers are given with respect to their stand mounting position. Note that the distance between the stand mounting and the centre of the driver along the x-axis and y-axis is 0\,cm and 11.5\,cm, respectively. Therefore, to align the listed loudspeaker coordinates to be with respect to the centre of the loudspeaker driver, rather than to the position of the stands, subtract 11.5\,cm from the given y-axis coordinates. Microphone height was measured from the floor to the centre of the array. Source and receiver heights were 150~cm and 155~cm, corresponding to the approximate average height of human mouth and ear height, respectively \cite{Roser2013}. For potential applications in which signals with varying amount of reverberation are important, the dataset was measured with five configurations of the variable acoustics room. The reverberation times for the five configurations \cite{Prawda2020} are presented in Table \ref{tab: reverb times}.

\begin{table}[htbp]
\caption{Coordinates of the source (S) and receiver (R) positions, corresponding to the numbering in Fig.\,\ref{fig:arni_roomplan}. To align the loudspeaker with respect to the centre of the loudspeaker driver, rather than with respect to the position of the stands, subtract 11.5\,cm from the given y-axis coordinates.}
\begin{center}
\begin{tabular}{|c|c|c|c|}
\hline
\textbf{Source/Receiver}&\multicolumn{3}{|c|}{\textbf{Position (m)}} \\
\cline{2-4} 
\textbf{Number} & \textbf{x}& \textbf{y}& \textbf{z} \\
\hline
S1 & 1.5 & 2.8 & 1.5 \\
S2 & 5.0 & 2.8 & 1.5 \\
S3 & 3.5 & 4.245 & 1.5 \\
\hline
R1 & 1.5 & 1.3 & 1.55 \\
R2 & 2.75 & 1.3 & 1.55 \\
R3 & 4.0 & 1.3 & 1.55 \\
R4 & 5.25 & 1.3 & 1.55 \\
R5 & 6 & 1.3 & 1.55 \\
R6 & 2.75 & 4.745 & 1.55 \\
R7 & 5.25 & 4.745 & 1.55 \\
\hline
\end{tabular}
\label{tab: coordinates}
\end{center}
\end{table}

\begin{table}[t]
\caption{Reverberation times of the variable acoustics room for the five absorption configurations.}
\begin{center}
\begin{tabular}{|c|c|c|c|}
\hline
\textbf{\% Absorbers}&\multicolumn{3}{|c|}{\textbf{Reverberation Time T$_{20}$(s)}} \\
\cline{2-4} 
\textbf{Enabled} & \textbf{{500~Hz}}& \textbf{{1000~Hz}}& \textbf{{2000~Hz}} \\
\hline
100 & 0.30 & 0.32 & 0.48 \\
75 & 0.38 & 0.40 & 0.56 \\
50 & 0.50 & 0.54 & 0.68 \\
25 & 0.72 & 0.76 & 0.84 \\
0 & 1.28 & 1.22 & 1.12 \\
\hline
\end{tabular}
\label{tab: reverb times}
\end{center}
\end{table}

\subsection{Specification}
\begin{figure}
		\captionsetup[subfigure]{justification=centering}
	\centering
	\begin{subfigure}[b]{\linewidth}\centering
		\includegraphics[trim=0 0 0 0,clip,width=\linewidth]{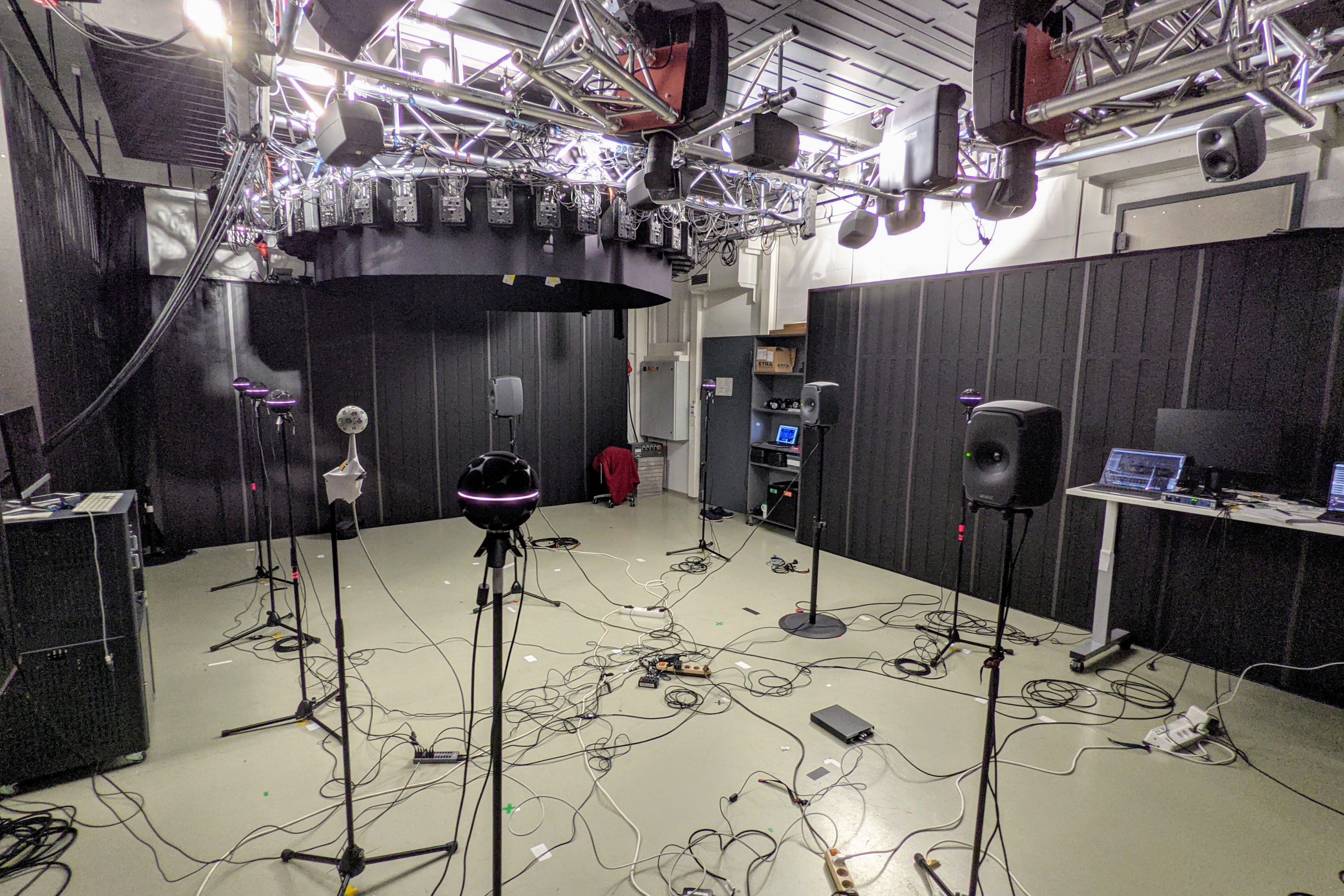}		
		\caption{Eigenmike SMA in position R2, 0\% absorbers enabled.} 
	\end{subfigure}
		 \par\medskip
	\begin{subfigure}[b]{\linewidth}\centering
		\includegraphics[trim=0 0 0 0,clip,width=\linewidth]{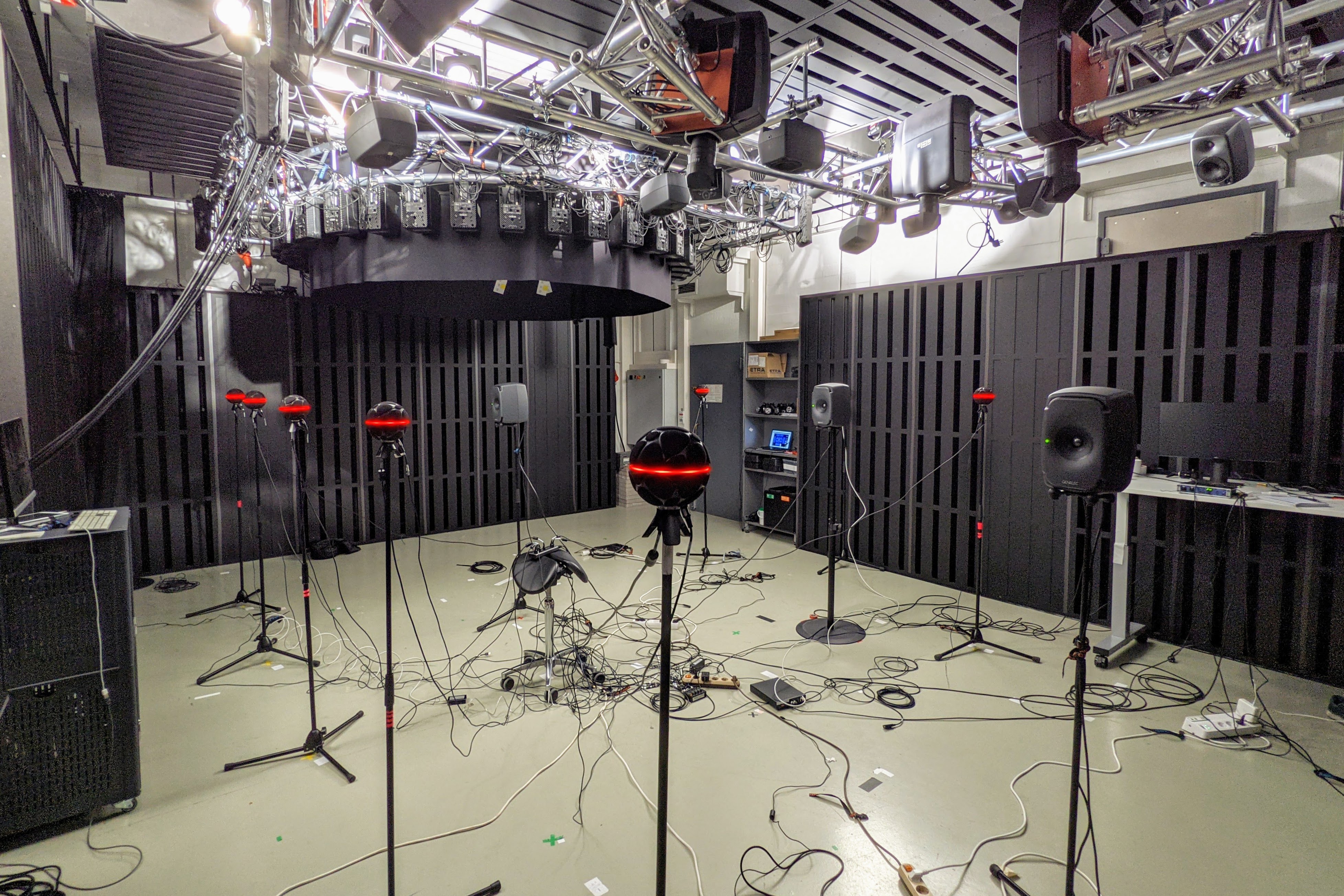}		
		\caption{Zylia SMA in recording state, 75\% absorbers enabled.} 
	\end{subfigure}
	\caption{Photographs of the loudspeakers and spherical microphone arrays with different absorption configurations.}
	\label{fig: arni photos}
\end{figure}

The dataset was captured with two different spherical microphone array (SMA) models: the mh Acoustics Eigenmike em32 and the Zylia ZM-1, which are 32 and 19 capsule SMAs capable of up to fourth-order and third-order Ambisonic capture, respectively. SRIRs were measured using 5 second exponential sine sweeps \cite{Farina2000}. Genelec 8331A coaxial loudspeakers were used for sweep playback at a peak A-weighted amplitude of approximately 80~dB, with an RME Fireface UCX audio interface. The Eigenmike SRIRs were recorded using the proprietary Firewire interface, and the seven Zylia ZM-1 microphones used the on-board USB interfaces. The Eigenmike SRIRs were recorded with the same SMA, which required moving to each receiver position separately, whereas the 7 Zylia SRIRs were recorded simultaneously using the Zylia 6DoF recording application. All audio was recorded at 24-bit resolution and 48\,kHz sample rate. Photographs of the recording setup with both the Eigenmike and the Zylia SMAs are presented in Fig.\,\ref{fig: arni photos}, with different variable acoustics absorption configurations.

Sweep recordings were repeated, and only the cleanest sweep was used. Impulse responses were generated through deconvolution of the measured sweeps with the inverse of the original sweep. The impulse responses were aligned temporally using onset detection based on the exceeding of a threshold magnitude, with a pre-onset gap of 5 samples (approximately 0.1~ms), and then delayed based on the geometric distances from the sources to receivers. Impulse responses were truncated to a length of 2.5\,s. 

\section{Encoding}

\begin{figure}
		\captionsetup[subfigure]{justification=centering}
	\centering
	\begin{subfigure}[b]{0.97\linewidth}\centering
		\includegraphics[trim=1.1cm 0 1.5cm 0.6cm,clip,width=\linewidth]{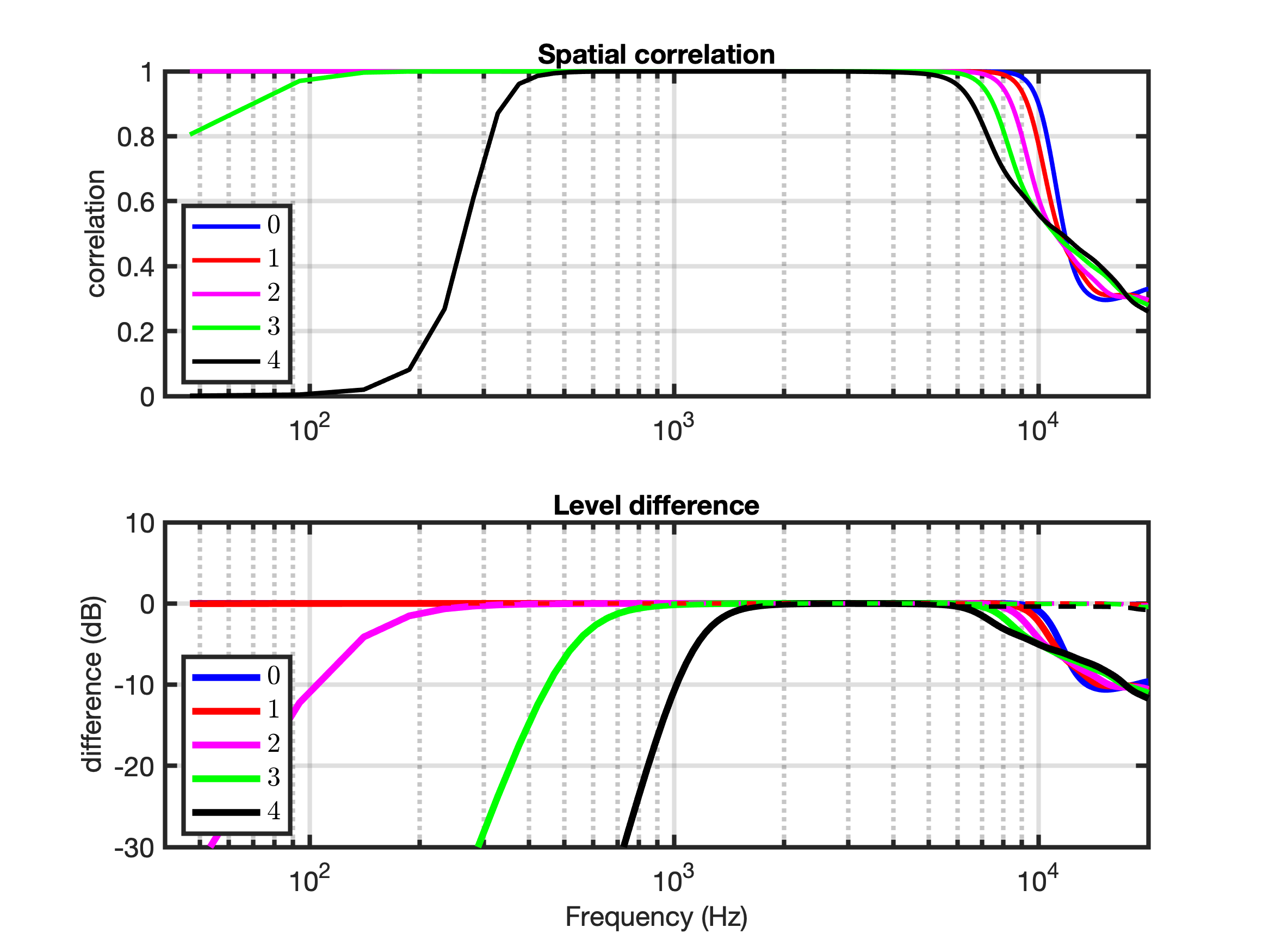}		
		\caption{Eigenmike em32 objective metrics results.} 
	\end{subfigure}
		 \par\medskip
	\begin{subfigure}[b]{0.97\linewidth}\centering
		\includegraphics[trim=1.1cm 0 1.5cm 0.6cm,clip,width=\linewidth]{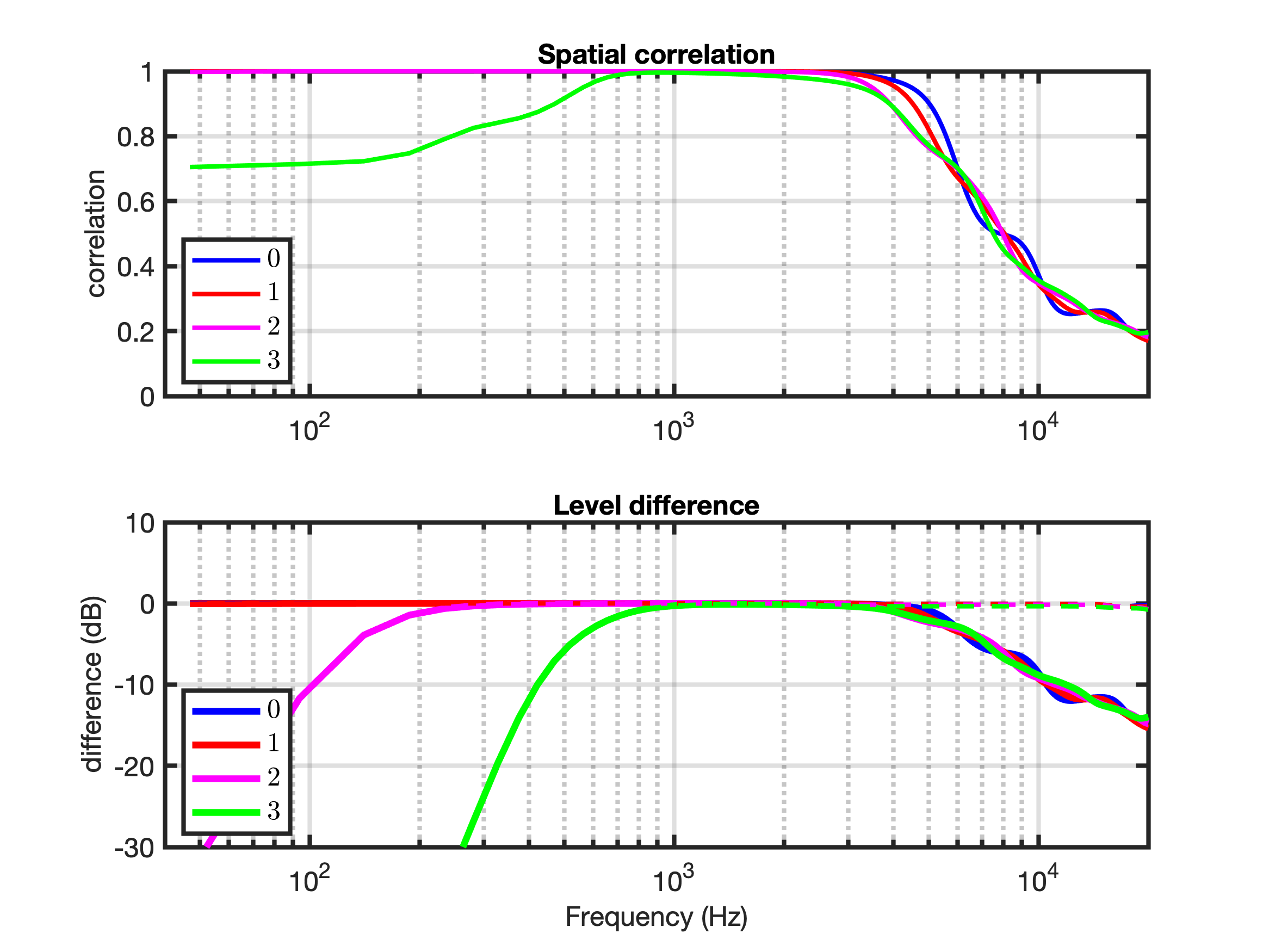}		
		\caption{Zylia ZM-1 objective metrics results.} 
	\end{subfigure}
	\caption{Objective metrics averaged over all components per SH order. The dotted lines in the level difference plots are after the patterns have been diffuse-field equalised above the array spatial aliasing frequency.}
	\label{fig: metrics}
\end{figure}

The raw SMA SRIRs were also converted to the spherical harmonic (SH) domain in third- and fourth-order Ambisonics using the open-source \texttt{array2sh} software tool described in \cite{mccormack2018real} for both SMAs. Tikhonov regularisation during encoding limited the gain to a set maximum $+15$\,dB. The encoded SRIRs follow the Ambisonic Channel Order (ACN) and semi-normalised (SN3D) conventions; often collectively referred to as the AmbiX format. Note that this encoder also applies diffuse-field equalisation of the encoded signals above the spatial aliasing frequency, as recommended by Gerzon in \cite{gerzon1975design}. This equalisation only affects the colouration of the signals, and not their spatial characteristics. Objective metrics, as described in \cite{moreau20063d,politis2017comparing}, are depicted in Fig.\,\ref{fig: metrics} for the Eigenmike and Zylia, which demonstrate the effect of this diffuse-field equalisation. Note that the theoretical spatial aliasing frequencies are approximately 3.3\,kHz and 5.1\,kHz 
for the Zylia and Eigenmike, respectively. The latter frequencies are also demonstrated in the spatial correlation metric as the point where the spatial correlation of the highest encoding order deviates from~1. However, it is also clear that the spatial aliasing frequency is order-dependent, and that the low-frequency cutoff for usable components (where the level difference metric deviates from 0\,dB, which is a product of the regularisation) is also order-dependent. Therefore, the approximate usable frequency ranges for each SH order, which narrow with increasing order, are provided in Table \ref{tab: usableHz} for convenience.

\begin{table}[t]
\caption{Approximate frequency bandwidths (Hz) where the SH encoded SRIRs exhibit the correct spatial patterns (spatial correlation of 1 and a level difference of 0\,dB with respect to ideal SH patterns), given the employed encoding approach. }
\begin{center}
\begin{tabular}{|c|c|c|c|c|}
\hline
\textbf{SH}&\multicolumn{2}{|c|}{\textbf{Eigenmike em32}}&\multicolumn{2}{|c|}{\textbf{Zylia ZM-1}}  \\
\cline{2-5} 
\textbf{Order} & lower & upper& lower & upper \\
\hline
0th & 50 & 8500 & 50 & 3500 \\
1st & 50 & 8000 & 50 & 3500 \\
2nd & 250 & 7000 & 250 & 3200 \\
3rd & 900 & 6000 & 900 & 3000 \\
4th & 1800 & 5100 & - & - \\
\hline
\end{tabular}
\label{tab: usableHz}
\end{center}
\end{table}

\section{Rotation correction for the Eigenmike}

Due to a poor fit of the Eigenmike to its shock mount, alignment of the array orientation was required. Here, the known direction of S1 from the perspective of receiver position R1 was used to derive a compensation rotation matrix to align a direction of arrival (DoA) estimate with this known true direction. The direct sound part of the (100\% absorbers enabled) SRIR was isolated, and the DoA found using the active-intensity vector \cite{fahy1990sound}, which is derived from a first-order encoding of the array SRIR. The required compensation rotation matrix $\mathbf{R} \in \mathbb{R}^{3 \times 3}$ was then calculated and used to create a SH domain rotation matrix $\mathbf{R}_\mathrm{SH} \in \mathbb{R}^{(N+1)^2 \times (N+1)^2}$ \cite{ivanic1998rotation}. The included SH domain conversions of the SRIR in the database have this compensation pre-applied. Since the relative Eigenmike orientations at different receiver positions were kept constant, this rotation offset is assumed to be appropriate for all 7 receiver positions. However, if using the raw/space-domain Eigenmike SRIRs for e.g. DoA estimation, then the Euler compensation rotation matrix should be used to correctly align them. The Zylia SRIR measurements do not require any rotation correction, since its forward facing direction is clearly indicated. 

\section{Dataset Download} \label{sec: dataset download}
The dataset is available under a Creative Commons Attribution 4.0 International license at \url{https://doi.org/10.5281/zenodo.5720724}, downloadable in Spatially Oriented Format for Acoustics (SOFA) \cite{Majdak2013} format. The dataset is available as both the raw capsule recordings and encoded SH format conversions, for both the Eigenmike em32 and Zylia ZM-1. Supplementary data includes the \textsc{Matlab} script and the appropriately windowed SRIR that were used for calculating the compensation rotation matrices for the Eigenmike, along with photographs and schematics of the source and receiver positions.

\bibliographystyle{IEEEtran}
\bibliography{ms.bib}

\begin{thebibliography}{10}
\providecommand{\url}[1]{#1}
\csname url@samestyle\endcsname
\providecommand{\newblock}{\relax}
\providecommand{\bibinfo}[2]{#2}
\providecommand{\BIBentrySTDinterwordspacing}{\spaceskip=0pt\relax}
\providecommand{\BIBentryALTinterwordstretchfactor}{4}
\providecommand{\BIBentryALTinterwordspacing}{\spaceskip=\fontdimen2\font plus
\BIBentryALTinterwordstretchfactor\fontdimen3\font minus
  \fontdimen4\font\relax}
\providecommand{\BIBforeignlanguage}[2]{{%
\expandafter\ifx\csname l@#1\endcsname\relax
\typeout{** WARNING: IEEEtran.bst: No hyphenation pattern has been}%
\typeout{** loaded for the language `#1'. Using the pattern for}%
\typeout{** the default language instead.}%
\else
\language=\csname l@#1\endcsname
\fi
#2}}
\providecommand{\BIBdecl}{\relax}
\BIBdecl

\bibitem{Roser2013}
M.~Roser, C.~Appel, and H.~Ritchie, ``Human height,''
  \url{https://ourworldindata.org/human-height}, 2013, accessed: 27/10/2021.

\bibitem{Prawda2020}
K.~Prawda, S.~J. Schlecht, and V.~V{\"{a}}lim{\"{a}}ki, ``{Evaluation of
  reverberation time models with variable acoustics},'' in \emph{Sound and
  Music Computing Conference}, 2020, pp. 145--152.

\bibitem{Farina2000}
A.~Farina, ``{Simultaneous measurement of impulse response and distortion with
  a swept-sine technique},'' in \emph{AES 108th Convention}, 2000.

\bibitem{mccormack2018real}
L.~McCormack, S.~Delikaris-Manias, A.~Farina, D.~Pinardi, and V.~Pulkki,
  ``Real-time conversion of sensor array signals into spherical harmonic
  signals with applications to spatially localized sub-band sound-field
  analysis,'' in \emph{AES 144th Convention}, 2018.

\bibitem{gerzon1975design}
M.~A. Gerzon, ``The design of precisely coincident microphone arrays for stereo
  and surround sound,'' in \emph{AES 50th Convention}, 1975.

\bibitem{moreau20063d}
S.~Moreau, J.~Daniel, and S.~Bertet, ``{3D} sound field recording with higher
  order ambisonics--objective measurements and validation of a 4th order
  spherical microphone,'' in \emph{AES 120th Convention}, 2006.

\bibitem{politis2017comparing}
A.~Politis and H.~Gamper, ``Comparing modeled and measurement-based spherical
  harmonic encoding filters for spherical microphone arrays,'' in \emph{IEEE
  Workshop on Applications of Signal Processing to Audio and Acoustics}, 2017,
  pp. 224--228.

\bibitem{fahy1990sound}
F.~J. Fahy and V.~Salmon, ``Sound intensity,'' \emph{Journal of the Acoustical
  Society of America}, vol.~88, no.~4, pp. 2044--2045, 1990.

\bibitem{ivanic1998rotation}
J.~Ivanic and K.~Ruedenberg, ``Rotation matrices for real spherical harmonics.
  direct determination by recursion,'' \emph{Journal of Physical Chemistry A},
  vol. 102, no.~45, pp. 9099--9100, 1998.

\bibitem{Majdak2013}
P.~Majdak, Y.~Iwaya, T.~Carpentier, R.~Nicol, M.~Parmentier, A.~Roginska,
  Y.~Suzuki, K.~Watanabe, H.~Wierstorf, H.~Ziegelwanger, and M.~Noisternig,
  ``{Spatially Oriented Format for Acoustics: a data exchange format
  representing head-related transfer functions},'' in \emph{AES 134th
  Convention}, 2013, pp. 1--11.

\end{thebibliography}

\end{document}